\documentclass[twocolumn]{emulateapj}

\begin{document}

\title{ Black Holes in Bulgeless Galaxies:\\
 An XMM-Newton Investigation of NGC 3367 and NGC 4536}

\author{W. McAlpine\altaffilmark{1}, S. Satyapal\altaffilmark{1}, M. Gliozzi\altaffilmark{1}, C. C. Cheung\altaffilmark{2}, R. M. Sambruna\altaffilmark{3}, Michael Eracleous\altaffilmark{4}}

\altaffiltext{1}{George Mason University, Department of Physics \& Astronomy, MS 3F3, 4400 University Drive, Fairfax, VA 22030}

\altaffiltext{2}{National Research Council Research Associate, National Academy of Sciences, Washington, DC 20001, resident at Naval Research Laboratory, Washington, DC 20375}

\altaffiltext{3}{NASA/GSFC, Code 662, Greenbelt, MD 20771}

\altaffiltext{4}{Department of Astronomy and Astrophysics and Center for Gravitational Wave Physics, The Pennsylvania State University, 525 Davey Lab, University Park, PA 16802}
    
\begin{abstract}

The vast majority of optically identified active galactic nuclei (AGNs) in the local Universe reside in host galaxies with prominent bulges, supporting the hypothesis that black hole formation and growth is fundamentally connected to the build-up of galaxy bulges. However, recent mid-infrared spectroscopic studies with {\it Spitzer} of a sample of optically ``normal'' late-type galaxies reveal remarkably the presence of high-ionization [NeV] lines in several sources, providing strong evidence for AGNs in these galaxies. We present follow-up X-ray observations recently obtained with {\it XMM-Newton} of two such sources, the late-type optically normal galaxies NGC 3367 and NGC 4536. Both sources are detected in our observations. Detailed spectral analysis reveals that for both galaxies, the 2-10 keV emission is dominated by a power law with an X-ray luminosity in the $L_{\rm 2-10 keV}$ $\sim$ 10$^{39}$ - 10$^{40}$ ergs s$^{-1}$ range, consistent with low luminosity AGNs. While there is a possibility that X-ray binaries account for some fraction of the observed X-ray luminosity, we argue that this fraction is negligible. These observations therefore add to the growing evidence that the fraction of late-type galaxies hosting AGNs is significantly underestimated using optical observations alone.  A comparison of the mid-infrared [NeV] luminosity and the X-ray luminosities suggests the presence of an additional highly absorbed X-ray source in both galaxies, and that the black hole masses are in the range of 10$^{5}$ - 10$^{7}$ $M_{\odot}$ for NGC 3367 and 10$^{4}$ - 10$^{6}$ $M_{\odot}$ for NGC 4536.
\end{abstract}


\section{Introduction}

The discovery that at the heart of virtually all early-type galaxies in the local Universe lies a massive nuclear black hole strongly suggests that black holes play a pivotal role in the formation and evolution of galaxies. The well-known correlation between the black hole mass, $M_{\rm BH}$, and the host galaxy stellar velocity dispersion $\sigma_\star$ \citep{gebhardt2000,ferrarese2000} suggests an intimate connection between black hole growth and the build-up of galaxy bulges, a connection that is reinforced by the fact that the vast majority of optically-identified active galactic nuclei (AGN) are found in early type hosts \citep[e.g.][henceforth H97]{heckman1980,ho1997}.

However, a number of recent studies have now shown that determining if AGNs reside in low-bulge galaxies cannot be definitively answered with only optical observations \citep[e.g.][]{satyapal2007,satyapal2008,satyapal2009,ghosh2008,desroches2009}. The problem arises because a putative AGN in a galaxy with a minimal bulge is likely to be energetically weak and deeply embedded in the center of a dusty late-type spiral. As a result, the optical emission lines can be dominated by the emission from star forming regions, severely limiting the diagnostic power of optical surveys in determining the incidence of AGNs in low-bulge systems. In a recent mid-infrared spectroscopic study with {\it Spitzer} of a sample of optically ``normal'' late-type galaxies, we found remarkably the presence of high-ionization [NeV] lines in a significant number of sources, providing strong evidence for AGNs in these galaxies, and suggesting that the AGN detection rate in late-type (Sbc or later) galaxies is possibly more than 4 times larger than what optical spectroscopic studies alone indicate \citep{satyapal2008}. While the detection of a [NeV] line strongly suggests that these galaxies harbor an AGN, [NeV] emission can originate in shock-heated gas produced by starburst-driven winds \citep{contini1997}. X-ray detection would provide corroborating evidence and strengthen the case for the presence of an AGN based on the luminosity and spectrum of the source.

X-ray observations arguably represent one of the most effective means to confirm the existence of an AGN and to investigate the AGN properties since: 1) Unlike the IR emission line features that probe lower density gas at larger distances from the black hole, X-rays are produced (and reprocessed) in the inner hottest nuclear regions where accretion occurs. 2) The penetrating power of (hard) X-rays allows them to carry information from the inner core of the galaxy without being significantly affected by absorption, and to probe the presence of the putative torus (provided that $N_{\rm H} < 10^{24}{~\rm cm^{-2}}$). 3) X-rays are far less affected by the host galaxy contamination than optical radiation. 4) The hard X-rays (2-10 keV) can more robustly constrain the bolometric luminosity of the AGN.

In this paper, we present a pilot study with {\it XMM-Newton} observations of NGC~3367 (D=43.6Mpc) \citep{tully1988} and NGC~4536 (D=17.7Mpc) \citep{saha2006}  which are two of the late-type galaxies in our {\it Spitzer} sample \citep{satyapal2008} showing robust evidence for a [NeV] line. The distance of NGC 4536 determined by use of Cepheids, is significantly higher than reported by \citet{tully1988}, yielding larger luminosities in all bands which are more consistent with an AGN. Among the galaxies with [NeV] detections, these 2 galaxies have optical spectra in the extreme HII-region range, indicating that there is absolutely no hint of an AGN based on their optical spectra. Both galaxies are isolated very low bulge systems; deep {\it HST} imaging of NGC~4536 for example, shows no evidence for a classical bulge but instead reveals a surface brightness profile consistent with a pseudo-bulge that exhibits spiral structure \citep{fisher2006a}. The presence of AGNs in these systems is indeed highly surprising.

\section{Observations and Data Reduction Procedure}

We observed our 2 targets with {\it XMM-Newton} in June, 2008. The nominal durations range between 27 and 34 ks (NGC 3367 - MOS: 34ks, PN: 32ks; NGC 4536 - MOS: 28ks, PN: 27ks). All of the EPIC cameras \citep{struder2001,turner2001} were operated in full-frame mode with the medium filter because of the presence of bright nearby sources in the field of view. As a precaution, for NGC 3367 and NGC 4536 the MOS cameras were operated in small window mode to prevent photon pile-up. The recorded events were screened to remove known hot pixels and other data flagged as bad; only data with {\tt FLAG=0} were used.  

The data were processed using the latest CCD gain values. For the temporal and spectral analysis, events corresponding to pattern 0--12 (singles, doubles, triples, and quadruples) in the MOS cameras and 0--4 (singles and doubles only, since the pn pixels are larger) in the pn camera were accepted. Arf and rmf files were created with the {\it XMM-Newton} Science Analysis Software (\verb+SAS+) 7.1. Investigation of the full--field light curves revealed the presence of several background flares for NGC 3367 and NGC 4536. These time intervals were excluded, reducing the effective total exposures time to the values reported in Table 1. The nuclear X-ray sources in NGC 3367 and NGC 4536 are well-separated from other field X-ray sources in the X-ray images (Figure 1) and extraction radii used for source spectra and light curves are 30\arcsec\ in both cases. For reference, 10\arcsec\ correspond to 2.11 kpc for NGC 3367, and 0.89 kpc for NGC 4536. The derived X-ray positions for the central sources are 10h46m35s +13d44m59s and 12h34m27s +02d11m16s for NGC 3367 and NGC 4536 respectively.  These positions are consistent with the positions of the optical nuclei taken from NED based on SDSS images (10h46m34.954s +13d45m03.09s and 12h34m27.050s +02d11m17.29s for NGC 3367 and NGC 4536 respectively). There are also off nuclear X-ray sources seen in Figure 1 which are not investigated in this paper. Background spectra and light curves were extracted from source-free circular regions on the same chip as the source, with extraction radii $\sim$2 times larger than those used for the source. There are no signs of pile-up in the pn or MOS cameras according to the {\tt SAS} task {\tt epatplot}. The RGS data  have signal-to-noise ratio ($S/N$) that is too low for a meaningful analysis.

The observation log with dates of the observations, EPIC net exposures, and the average count rates are reported in Table 1. 

The spectral analysis  was performed using the {\tt XSPEC v.12.3.1} software package \citep{arnaud1996,dorman2001}. The EPIC data have been re-binned in order to contain at least 20 counts per channel, depending on the brightness of the source. The errors on spectral parameters are at 90\% confidence level for one interesting parameter ($\Delta \chi^2=2.71$).

\begin{table}
\begin{center}
\scriptsize
\begin{tabular}{lccc}
\multicolumn{4}{l}{{\bf Table 1: Observation Log}} \\
\hline
\hline
\noalign{\smallskip}
Source & Date & PN exposure & PN rate \\    
\noalign{\smallskip}
 & [dd/mm/yyyy] & [ks] & [${\rm s^{-1}}$] \\
\noalign{\smallskip}
\hline
\noalign{\smallskip}
 NGC~3367 & 06/16/2008 & 17.8 & $0.100\pm0.003$ \\
\noalign{\smallskip}
\hline
\noalign{\smallskip}
 NGC~4536 & 06/17/2008 & 14.4 & $0.008\pm0.003$ \\ 
\noalign{\smallskip}
\hline
\hline
\end{tabular}
\end{center}
\footnotesize 
\label{tab1}
\end{table}

\begin{figure}[]
\plotone{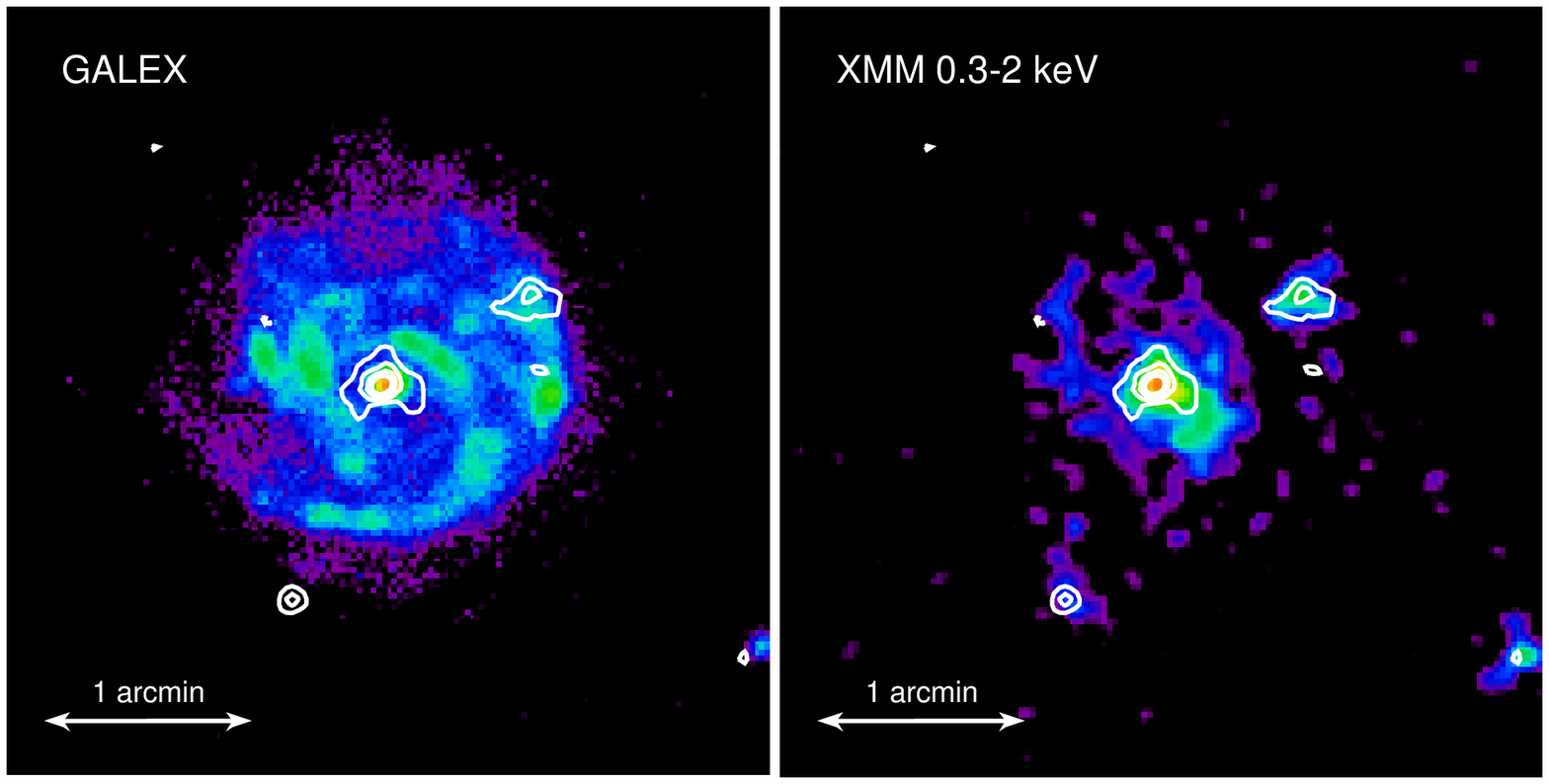}
\plotone{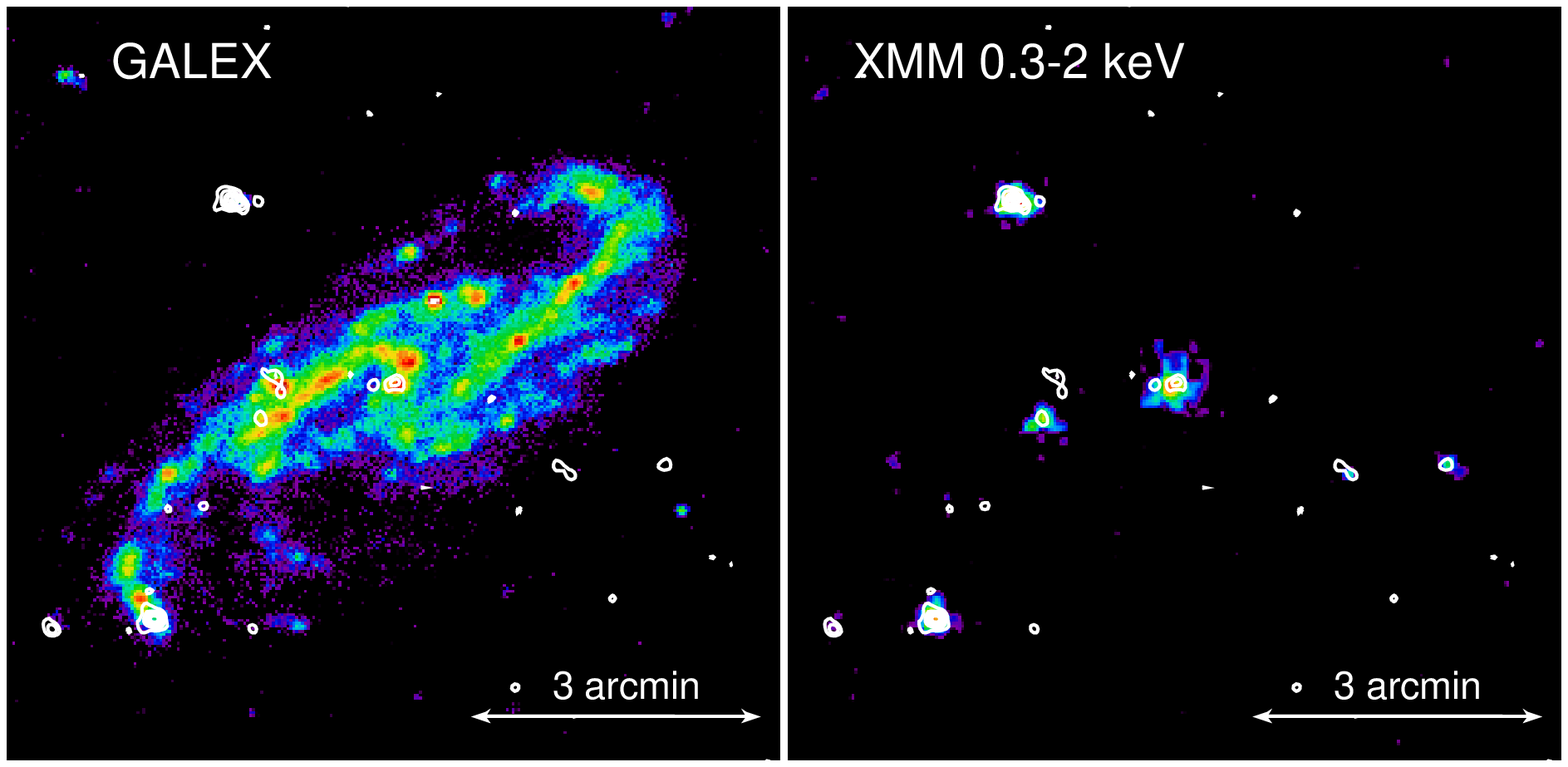}
\caption{ Images of NGC 3367 (top) and NGC 4536 (bottom) from GALEX in the near-ultraviolet (left; from \citet{gil2007}) and smoothed XMM in the 0.3-2 keV energy range (right). Both panels show the smoothed XMM 2-10 keV image overlaid as contours with levels starting at 0.25 counts (smoothed), incremented in 0.25 steps. The smoothing of the XMM images used a Gaussian with kernel of radius = 4.8" (3 pixels).}
\end{figure}

\section{Results}

\begin{table*}
\begin{center}
\scriptsize
\begin{tabular}{lccccccc}
\multicolumn{8}{l}{{\bf Table 2: Best Fit Model Results for PN Data }} \\
\hline
\hline
\noalign{\smallskip}
 Name &Power Law &Thermal &Intrinsic &0.3-2 keV &2-10 keV &$\%$L$_{\rm PL}$ & $ \chi^2_{red}$(dof)\\    
 & Photon Index &  Component  kT & N$_{\rm H}$ & Luminosity & Luminosity & 2-10 keV& \\
\noalign{\smallskip}
 & &(keV)&$10^{21}$cm$^{-2}$&$10^{40}$ ergs s$^{-1}$&$10^{40}$ ergs s$^{-1}$& & \\
\noalign{\smallskip}
\hline
\noalign{\smallskip}
 NGC 3367 & $2.15\pm0.08$ & $0.64\pm0.03$ & ----- & 3.3  & 2.0 &98&1.34(88)  \\
\noalign{\smallskip}
\hline
\noalign{\smallskip}
 NGC 4536 & $2.3\pm0.3$ & $0.58\pm0.03$ & $1.1\pm0.4$ & 1.7  & 0.5 &98&1.23(69)   \\
\noalign{\smallskip}
\hline
\hline
\end{tabular}
\end{center}
\footnotesize
\label{tab2}
\end{table*}

\begin{figure*}[]
{\includegraphics[bb=40 2 567 704,clip=,angle=-90,width=9.cm]{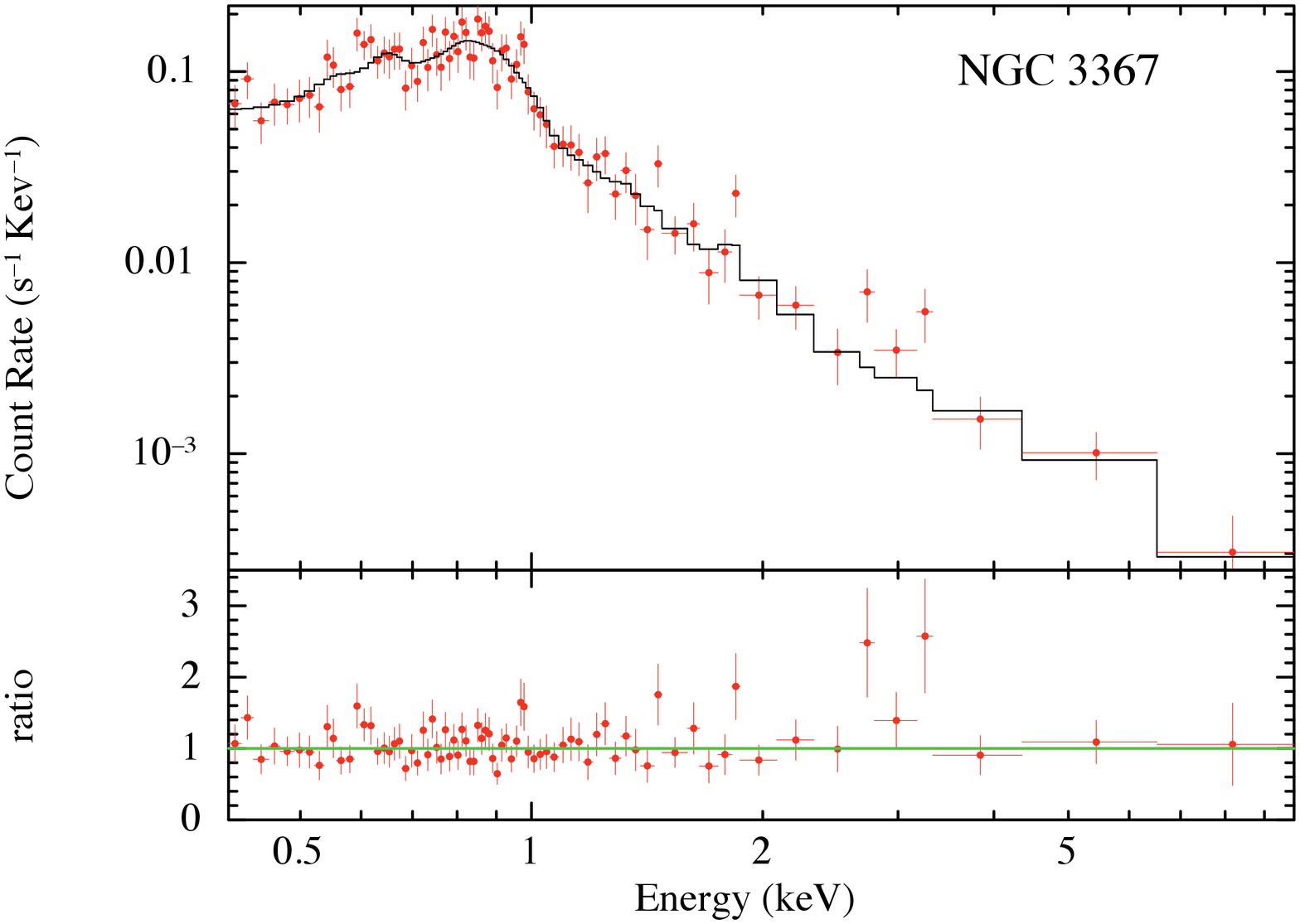}}
{\includegraphics[bb=40 2 567 704,clip=,angle=-90,width=9.cm]{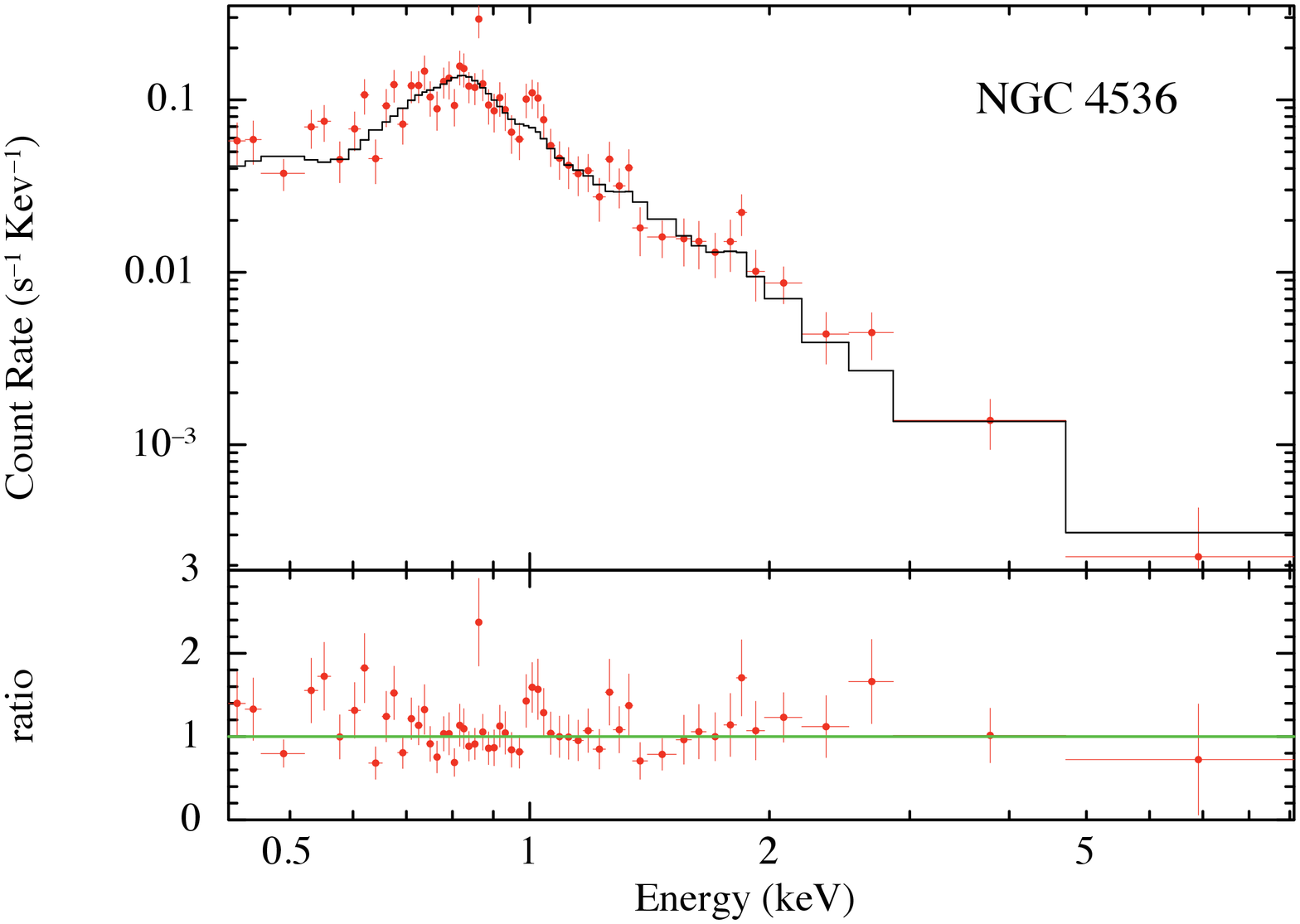}}
\caption{Best fit models and data-to-model ratios obtained by fitting the EPIC pn spectra in the 0.3-10 keV energy range. Both models include a best-fit power law plus a thermal component and are absorbed by Galactic N$_{H}$. NGC 4536 requires additional intrinsic absorption.}
\end{figure*}

\subsection{X-Ray Spectral Analysis}

We fitted the EPIC spectral data in the 0.3 - 10 keV range. To analyze the spectral data we first attempted to find the best-fit model for the pn data, which would then be applied to the MOS1 and MOS2 data. For NGC 3367 we fitted the data with a simple power law with galactic absorption which did not provide an acceptable fit. Positive residuals at low energy suggest the addition of a thermal component which provided an acceptable fit with a $ \chi^2_{red}=1.34$ for 88 degrees of freedom. Allowing the column density to vary or adding additional absorption components did not improve the fit. Therefore we determined that the best-fit model shown in Figure 2 was produced by a power law with an added thermal component. The model provided a $kT$ value of $0.64\pm0.03$ keV for the thermal component and a photon index value of $2.15\pm0.08$. Applying this model to the pn, MOS1, and MOS2 data resulted in a good fit with a $\chi^2_{red}=1.25$ for 140 degrees of freedom. A summary of the results is shown in Table 2.

We then determined the contribution of the components to the overall flux by evaluating each component separately. Removing the power-law component and leaving the thermal component gave fluxes of $6.3\times 10^{-14}$ and $1.7 \times 10^{-15}$ ergs cm$^{-2}$ s$^{-1}$ for the 0.3--2 keV and the 2--10 keV ranges respectively. Restoring the power-law component and removing the thermal component gave fluxes of $6.0\times 10^{-14}$ and $9.1\times 10^{-14}$ ergs cm$^{-2}$ s$^{-1}$ for the 0.3--2 keV and the 2--10 keV ranges respectively. From these values it can be determined that the power-law component dominates in the 2--10 keV energy range with 98$\%$ of the luminosity.

For NGC 4536 we again fitted the data with a simple power law with galactic absorption which also did not result in an acceptable fit. As before, we then added a thermal component, which resulted in an acceptable fit with a $ \chi^2_{red}=1.37$ for 70 degrees of freedom. Next we added an absorber at the position of the source, which further improved the fit to a $ \chi^2_{red}=1.23$ for 69 degrees of freedom. The best-fit model included a thermal component with a $kT$ value of $0.58\pm0.03$ keV, a photon index value of $2.30\pm0.26$, and an absorber with $N_{\rm H} = (1.14 \pm 0.4) \times 10^{21}$ cm$^{-2}$. Applying this model to the pn, MOS1, and MOS2 data resulted in a reasonably good fit with a $ \chi^2_{red}=1.24$ for 106 degrees of freedom.  

Lastly we determined the contribution of the different components to the overall flux. Removing the power-law component and leaving the thermal component gave fluxes of $5.5\times 10^{-14}$ and $1.7 \times 10^{-15}$ ergs cm$^{-2}$ s$^{-1}$ for the 0.3--2 keV and the 2--10 keV ranges respectively. Restoring the power-law component and removing the thermal component gave fluxes of $6.1 \times 10^{-14}$ and $6.0\times 10^{-14}$ ergs cm$^{-2}$ s$^{-1}$ for the 0.3--2 keV and the 2--10 keV ranges respectively. From these values we again determined that the power-law component dominates in the 2--10 keV range with 98$\%$ of the luminosity.

\subsection{X-Ray Temporal Analysis}

 Since in general the X-ray variability is one of the defining properties of AGNs, it is important to investigate the temporal properties of NGC 3367 and NGC 4536. We studied the short-term variability of both sources using EPIC pn data with time-bins of 1000 s. Figure 3 shows the EPIC time series in the 0.3-10 keV band. A visual inspection of Figure 3 reveals that the average count rate of NGC 3367 is smaller than the one of NGC 4536 by a factor of $\sim$10, and suggests that low-amplitude flux changes might be present in both light curves. However, a formal analysis based on a  $\chi^2$ test indicates that there is statistically significant variability only in NGC 4536 ($P_{\chi^2}\simeq 9\%$), while no significant variability ($P_{\chi^2}\simeq 83\%$) is detected in NGC 3367.

\begin{figure*}[]
{\includegraphics[bb=90 60 550 320,clip=,angle=0,width=9.cm]{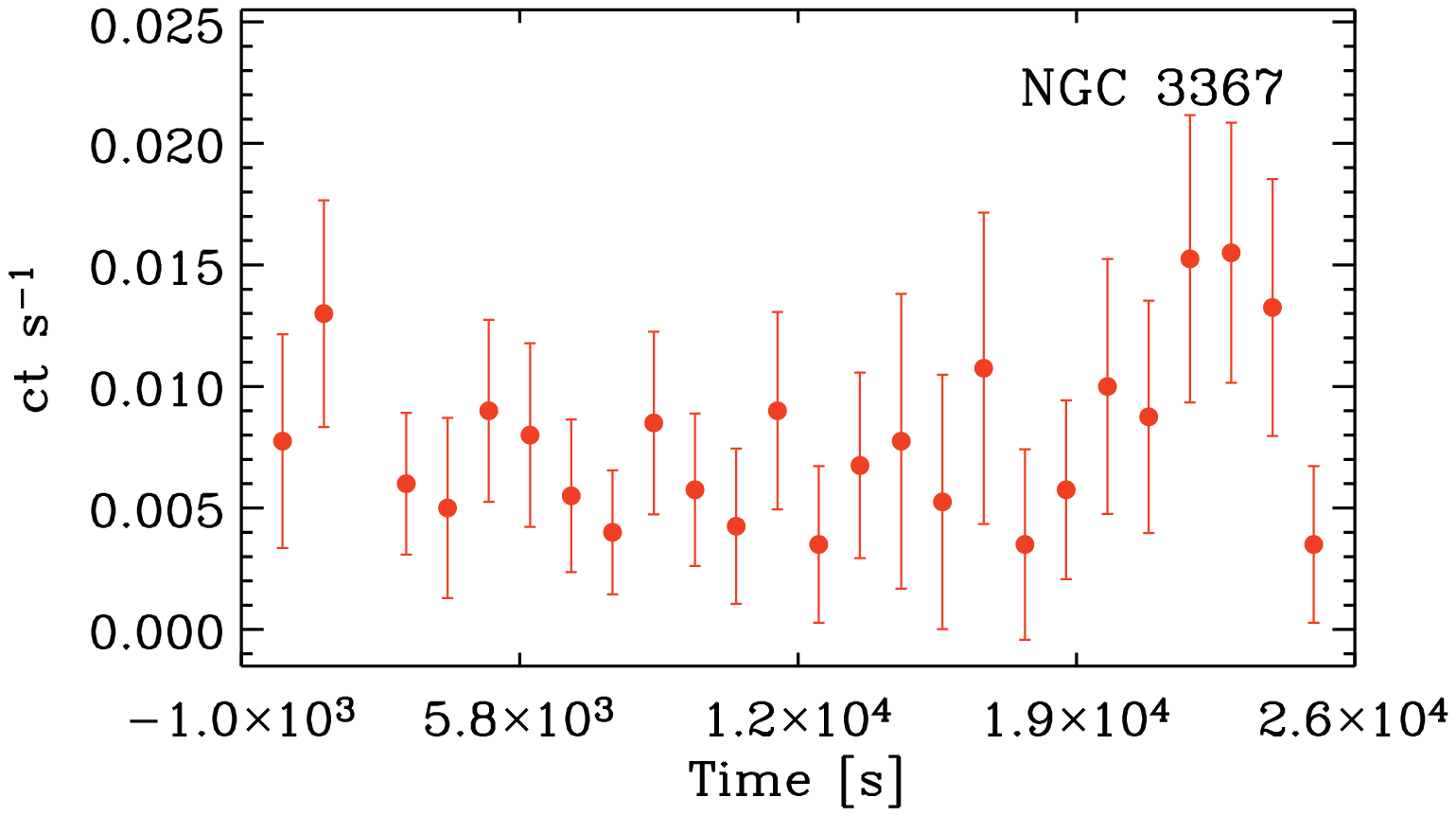}}
{\includegraphics[bb=90 60 550 320,clip=,angle=0,width=9.cm]{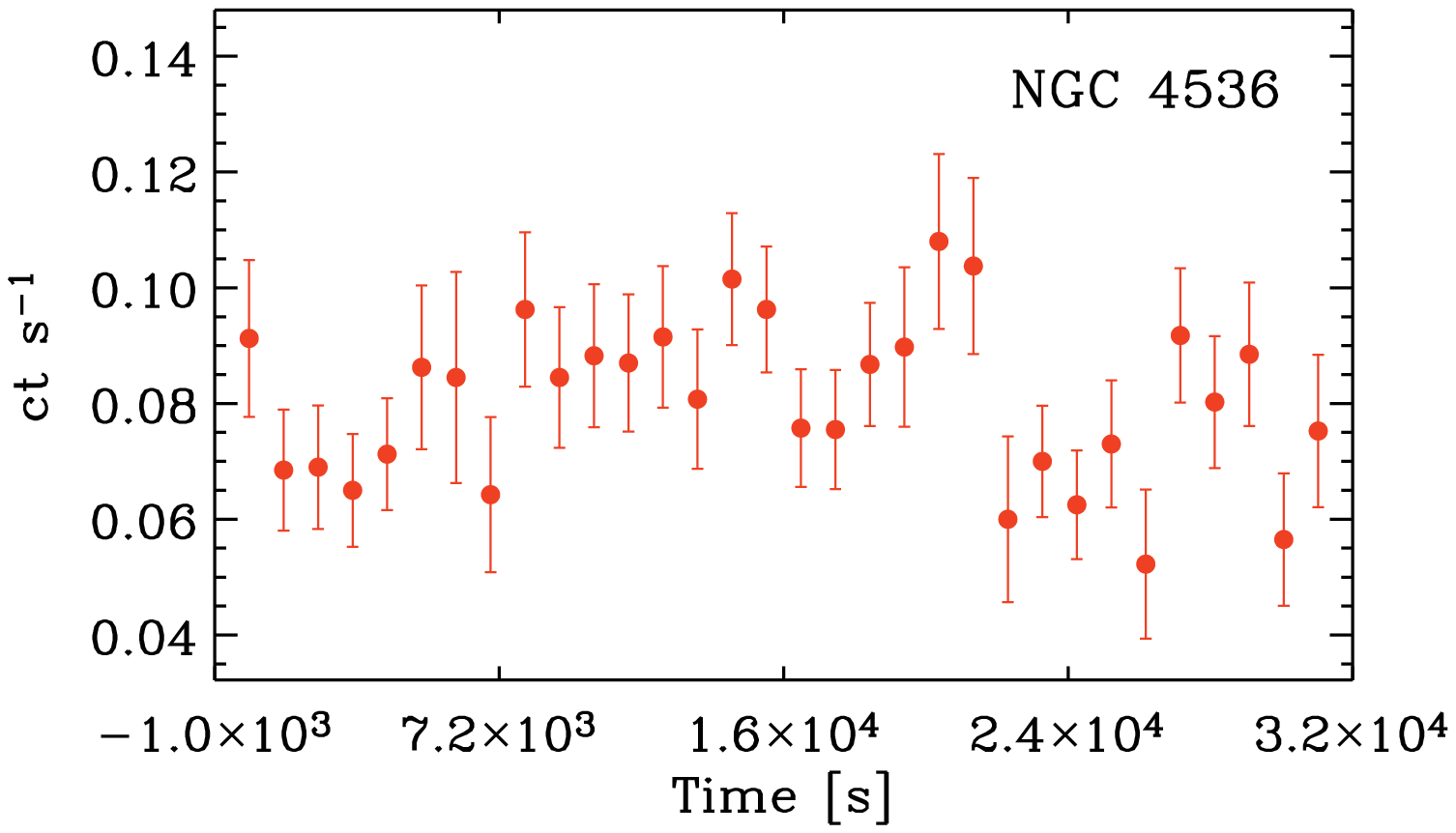}}
\caption{{\it XMM-Newton} EPIC light curves for NGC 3367 and NGC 4536. The time
bin is 1000 s.}
\end{figure*}

\subsection{UV - Optical Data}

The OM allows us to simultaneously investigate the optical-UV properties of NGC 3367 and NGC 4536. In the case of these galaxies, the AGN is of low luminosity and furthermore obscured in the optical. It is thus likely that the optical-UV emission is dominated by star formation in the host galaxy in addition to being attenuated by dust. Nonetheless, we compare the UV-X-ray properties of NGC 3367 and NGC 4536 to optically identified more powerful AGNs to see if they are similar. In particular, we can compare the broadband spectral index $\alpha_{OX}$ between our galaxies and optically identified AGNs. Using a sample of 333 optically selected Seyferts, \citet{steffen2006} found that the broadband spectral index $\alpha_{OX}$ are highly correlated with the UV monochromatic luminosity. For our sample, we apply the same approach by computing the spectral index, $\alpha_{OX}$ = log($l_{2500 \AA} / l_{2 keV}$)/ log($\nu_{2500 \AA} / \nu_{2 keV}$) \citep{tananbaum1979}, where $l_{\nu}$ is the monochromatic luminosity in units of ${\rm erg~s^{-1}~Hz^{-1}}$.  This is then plotted against the UV monochromatic luminosity. The 2500 \AA\ flux is obtained by converting the flux measured in the UVM2 band (2310 \AA) assuming a typical slope of 0.7 ($f_{\nu} \propto \nu^{-0.7}$). Different extinction prescriptions were tested to account for the reddening in the UV band; the Small Magellanic Cloud extinction law was finally used for NGC 3367 and NGC 4536 \citep[see][for details]{gliozzi2008}. E(B-V) values of 0.029 mag and 0.018 mag were used for NGC 3367 and NGC 4536 respectively.

The observed magnitudes for NGC 3367 and NGC 4536 are 14.3 and 15.2 respectively.  The extinction corrected fluxes in the UVM2 band are 1935 $\mu$Jy and 770 $\mu$Jy for NGC 3367 and NGC 4536 respectively.  The fluxes at 2500 \AA\ are 2046 $\mu$Jy and 814 $\mu$Jy for NGC 3367 and NGC 4536 respectively.

The values of $\alpha_{OX}$ for our sample are shown superimposed to the best-fit linear regression found by \citet{steffen2006} in Figure 4. As can be seen from the figure, NGC 3367 and NGC 4536 are clear outliers to the relation and are underluminous in the X-ray as compared to the sample of Seyferts. A possible explanation for this discrepancy may be that optical and UV emission is dominated by star formation.  This is reinforced by the fact that these
objects are optically classified as H II galaxies, suggestive of nuclear star formation.

\begin{figure}[]
\plotone{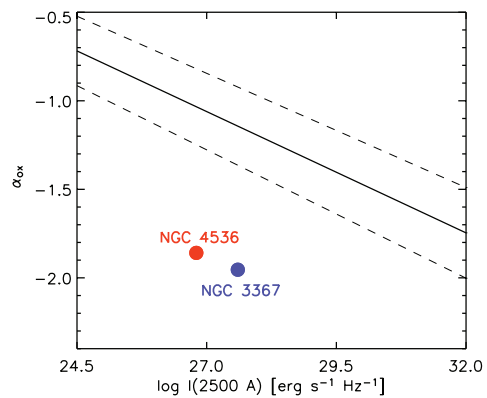}
\caption{$\alpha_{OX}$ plotted versus 2500 \AA ~monochromatic luminosity. The solid line corresponds to the best-fit relation derived by Steffen et al. (2006) and the dashed lines account for the uncertainties in the fit. }
\end{figure}

\section{Previous Observations of NGC 3367 and NGC 4536}

NGC  3367 is an isolated face-on Sc barred galaxy \citep{devaucouleurs1976} that is optically classified as an HII galaxy (H97). Although optically there is no hint of an AGN, radio observations reveal a bipolar synchrotron outflow from an unresolved compact nucleus with a diameter $<$65 pc \citep{garcia2002}, and the possibility of two lobes straddling the nucleus and extending up to 12 kpc \citep{garcia1998}. Assuming a spectral index in the range of, $\alpha=0-0.7$, we estimate a 5 GHz flux for the nucleus of 1.0 to 1.4 mJy based on the VLA 8.4 GHz measurement of 0.96 mJy at $\lesssim$ 0.3\arcsec\ resolution by \citet{garcia2002}. The higher spectral index assumed is that measured for the core at matched lower resolution (4.5\arcsec) between 1.4 GHz (as published in \citet{garcia1998} and 8.5 GHz, the latter being newly determined from our analysis of an archival VLA dataset (program AK550). However this spectral measurement may be dominated by diffuse emission surrounding the core as seen in the maps from \citet{garcia1998}.  In a study of M81, another low luminosity AGN, $\alpha$ = 0-0.3 has been observed \citep{markoff2008}. The central stellar velocity dispersion is  61.2 $\pm$ 10.1 km s$^{-1}$ \citep{ho2009}. 

In the MIR {\it Spitzer} observation of NGC 3367 the [NeV] lines at 14.3 $\mu$m and 24.3 $\mu$m were detected \citep{satyapal2008}. The detected fluxes of these lines were 1.2 $\pm$ 0.3 and 0.93 $\pm$ 0.24  10$^{-22}$W cm$^{-2}$ respectively. The aperture from which the [NeV] 14.3 µm line was detected corresponds to a projected size of $\sim$1 $\times$ 2.4 kpc. Photoionization models of mid-infrared fine-structure line fluxes suggests that about 10 to 30 percent of the total bolometric luminosity of the galaxy may attributed to the AGN.

NGC 4536 is a barred late-type spiral SABbc \citep{devaucouleurs1976} optically classified as an HII galaxy (H97). {\it HST} imaging shows no evidence of a classical bulge, but has a surface brightness profile consistent with a pseudo-bulge that appears to exhibit spiral structure \citep{fisher2006b}. To the best of our knowledge, the only hint of an AGN in the optical is the optical line ratios from Space Telescope Imaging Spectrograph spectroscopy that suggest the presence of a weak AGN \citep{hughes2005}. There is evidence of star formation in the central $\sim$ 20 \arcsec\ $\times$ 30 \arcsec\ region with strong Br$\gamma$ emission \citep{puxley1988} and 10.8 $\mu$m emission \citep{telesco1993}. Also a continuum-subtracted image of H$\alpha$ + [NII] emission \citep{pogge1989} is suggestive of nuclear star formation. The radio emission has a diffuse morphology with three separate peaks \citep{vila1990,laine2006} that may be an annular ring of star formation around the nucleus. The morphology is similar to what is seen at 10.8 $\mu$m \citep{telesco1993}, as well as in the 1-0 S(1) molecular hydrogen line \citep{davies1997}. ROSAT High Resolution Imager data revealed two ultra luminous X-ray sources, one of which may be coincident with the optical nucleus \citep{liu2005}. The second ULX is 150\arcsec\ away from the nucleus of the galaxy, well outside the 30\arcsec\ aperture radius of the {\it XMM-Newton} observation. The central region of NGC4536 is composed of an extended radio source at arcsecond resolution \citep{vila1990}. From our reanalysis of the VLA 4.9 GHz map (1.3$"$ beam; program AD176) published by \citet{vila1990}, we measure a point source upper limit of $<$1.7 mJy for the nucleus based on the observed peak surface brightness at the optical center. The central stellar velocity dispersion is 85 $\pm$ 1 km s$^{-1}$ \citep{batcheldor2005}.  

The [NeV] 14.3 $\mu$m line was detected in this galaxy from an $\sim$770 $\times$ 770 pc region, approximately 10\arcsec\ northeast of the optical nucleus \citep{satyapal2008}. A flux of 0.32 $\pm$ 0.09 $\times$ 10$^{-21}$ W cm$^{-2}$ was detected. [OIV] 26 $\mu$m and [NeIII] 15.5 $\mu$m lines were also detected, but were concentrated at a location nearer to the optical nucleus than the [NeV] emission. Photoionization models of mid-infrared fine-structure line fluxes suggests that about 10 to 30 percent of the total bolometric luminosity of the galaxy may attributed to the AGN.

\section{Contamination by X-Ray Binaries}

The 2-10 keV band X-ray luminosity of 2.0 x $10^{40}$ ergs s$^{-1}$ and 0.9 x $10^{40}$ ergs s$^{-1}$ for NGC 3367 and NGC 4536 respectively is consistent with low-luminosity AGNs \citep{ho2001}. A compilation of low-luminosity AGN from \citet{ho2001}, finds X-ray luminosities ranging from $\sim 10^{37} - \sim 10^{41}$ ergs s$^{-2}$

While the observed X-ray emission is most likely due to the presence of an AGN, there is a possibility of contamination due to X-ray binaries (XRBs). \citet{gilfanov2004} examined the connection between the X-ray luminosity of low mass X-ray binaries (LMXBs) and the stellar mass of 11 galaxies with no evidence of current star formation. An average ratio of $L_{\rm x}/M_{\star}$ = 8.3 $\times$ 10$^{28}$ ergs s$^{-1}$ $M_{\odot}^{-1}$ was found for the 0.3-10 keV luminosity. For NGC 3367 and NGC 4536, LMXBs would account for approximately only 5 percent of the observed X-ray luminosity according to this ratio.  

High mass X-ray binaries (HMXBs) however may account for significantly higher portion of the observed luminosity since both galaxies show emission lines consistent with ongoing star formation. \citet{colbert2004} used a sample of 32 spiral, elliptical, and irregular galaxies, with 1441 X-ray point sources detected by Chandra to study the relation between the X-ray emission (0.3-8 keV) from these point sources and the properties of the galaxy. They related the X-ray point source luminosity, assumed to be mostly from black hole HMXBs and ULXs, to K-band luminosity and the FIR-UV luminosity. Using the K-band luminosity to determine the stellar mass and the FIR-UV luminosity to determine the star formation rate (SFR) they found a linear relationship between X-ray luminosity, stellar mass, and the SFR of a galaxy. Using this relation, we estimated the luminosity expected from X-ray sources other than an AGN. SFRs for NGC 3367 and NGC 4536 were estimated using the extinction-corrected UV luminosities calculated earlier, and using the formula given in \citet{kennicutt1998}. With SFRs of 0.65 and 0.16 M$_{\odot}$ yr$^{-1}$ for NGC 3367 and NGC 4536 respectively, the relation gives an X-ray luminosity from point sources of approximately 10 percent of the observed luminosity from both NGC 3367 and NGC 4536, suggesting that AGN dominates the X-ray luminosity.

The possible contamination by XRBs may also be estimated based on their expected density within the area observed by the 30\arcsec\ beam of the {\it XMM-Newton}.  \citet{pence2001} examined the distribution of X-ray point sources within NGC 5457. Using the surface density of X-ray point sources in NGC 5457, a rough estimate of the number of X-ray sources in NGC 3367 and NGC 4536 may be determined. They determine the surface density of X-ray sources as a function of the distance from the nucleus. Assuming a distance of 6.85 Mpc for NGC 5457 \citep{saha2006}, the distance from the nucleus may be compared with our galaxies. The {\it XMM-Newton} extraction aperture for NGC 3367 corresponds to a radius of 6.3 kpc and is estimated to have a surface density of 0.29 sources per kpc$^{2}$. The {\it XMM-Newton} extraction aperture for NGC 4536 corresponds to a radius of 2.6 kpc and is estimated to have a surface density of 0.56 sources per kpc$^{2}$. These values give an expected number of sources of approximately 36 and 12 sources for NGC 3367 and NGC 4536 respectively. Typical XRBs have luminosities of up to 10$^{37}$ ergs s$^{-1}$ \citep{white1995}. However, assuming all the X-ray sources in NGC 3367 and NGC 4536 emit at $\sim$10$^{38}$ ergs s$^{-1}$ (the highest luminosities reached by XRBs in NGC 5457, \citet{pence2001} NGC 3367 would require at least 200 sources and NGC 4536 would require at least 90 sources in order to account for the observed X-ray emission in the 2-10 keV band if these sources were exclusively due to XRBs. While we assumed a similarity between NGC 5457 and our sample that may not exist, the probability of these many sources emitting at an average luminosity of 10$^{38}$ ergs s$^{-1}$ within an area of a few kpc$^{2}$ is extremely unlikely.

In addition to the lower X-ray luminosities, XRBs are expected to be characterized by different SEDs over the UV - optical - X-ray range compared to those from AGNs \citep{yuan2005}. By using the [NeV] emission as a proxy for UV emission, the ratio of the [NeV] luminosity to the X-ray luminosity can be used as a crude indicator of the SED of the ionizing source. The L[NeV]/Lx ratio for NGC 3367 and NGC 4536 is 0.130 and 0.013 respectively, consistent with optically identified AGN \citep{gliozzi2009}. The $L_{\rm [NeV]}/L_{\rm x}$ ratio will of course depend on a number of parameters including the ionization parameter and the fraction of the emission due to star formation. Moreover there is a large scatter in the correlation between [NeV] emission and UV emission. However, the similarity between this ratio in NGC 3367 and NGC 4536 suggests that XRBs alone cannot be responsible for both the X-ray and [NeV] line emission in NGC 3367 and NGC 4536. Lastly, the X-ray variability presented in section 3.2 for NGC 4536 strongly suggests that the emission is from a single source.

We have thus shown that the X-ray luminosities of NGC 3367 and NGC 4536 are unlikely to be due to XRBs. Furthermore, the detection of the [NeV] line together with the high X-ray luminosities, is highly suggestive of an AGN.

\section{Bolometric Luminosities}

An estimation of the bolometric luminosity of the AGN may be found using the X-ray and [NeV] luminosities using their respective bolometric correction factors. If we assume that the X-rays are emitted exclusively by the AGN, the X-ray luminosity can be used to estimate the bolometric luminosity of the AGN. If we assume that the black holes at the center of these galaxies have relatively low masses given their low bulge masses, then the Eddington ratios of these galaxies are likely to be high \citep{greene2007}. \citet{lusso2010} studied 545 X-ray selected type 1 AGNs from the XMM-COSMOS survey with available black hole masses and bolometric luminosities estimated using their spectral energy distribution. Assuming that these AGNs emit at very high rates with Eddington ratios $>$0.2 we adopt a mean bolometric correction factor of 53 \citep{lusso2010}. Using this bolometric correction factor, the estimated bolometric luminosities are log$L_{\rm bol}$(ergs s$^{-1}$) = 42.0 and log$L_{\rm bol}$(ergs s$^{-1}$) = 41.7 for NGC 3367 and NGC 4536, respectively.

For these galaxies photoionization by a starburst is unlikely to give rise to significant [NeV] emission \citep{abel2008}. If we assume that the [NeV] only arises from the AGN, we can use the relation between bolometric luminosity and [NeV] luminosity established in a large sample of more powerful AGNs \citep{satyapal2007} to estimate the bolometric luminosity of the AGNs. The bolometric luminosities are estimated to be log$L_{\rm bol}$(ergs s$^{-1}$) = 43.3 and log$L_{\rm bol}$(ergs s$^{-1}$) = 42.0 for NGC 3367 and NGC 4536, respectively. These luminosities are a factor of 20 and 2 higher, respectively, than the bolometric luminosities estimated using the X-ray luminosities. This discrepancy may suggest that some of the [NeV] emission originates in shocks.  However, given the large scatter in the correlations used to estimate the bolometric luminosities, any definitive conclusions cannot be made about the discrepancy.

\section{Black Hole Mass Estimation Methods}

{\it Eddington Mass Limits} --- Using the [NeV] and X-ray estimated bolometric luminosities, we can obtain lower limits on the mass of the black hole assuming the Eddington limit.  The black hole mass estimates are given in Table 3. 

{\it Bulk Motion Comptonization (BMC) Model} --- One possible method to determine the mass in BH systems relies on the fact that hard X-rays are produced by the Comptonization process in both stellar and supermassive black holes. Specifically, by fitting the X-ray spectra of Galactic black holes (GBHs) with the BMC model during their spectral transition, one  can derive  a universal scalable relationship between the photon index $\Gamma $ and the normalization of the BMC model $N_{\rm BMC}$. If one assumes that the physics of a black hole system is the same no matter the scale and that the bulk of the X-ray emission is produced by Comptonization, this technique can be used to compare masses of black holes of differing scales. Using the known mass and properties of a Galactic black hole used as reference, the mass of the black hole in question can be extrapolated. A more detailed description of the application of this method to AGNs is given in \citet{gliozzi2009,gliozzi2010}. The mass estimates obtained using this technique vary depending on the uncertainty of the BMC spectral parameters and on the GBH used as a reference giving a range of values. This mass range for NGC3367 and the upper limit mass for NGC4536 is compatible with the lower mass limit derived using the Eddington limit as can be seen in Table 3.

{\it M-$\sigma$ Relation} --- The mass of the central black holes can be estimated using the M-$\sigma$ relation, assuming that the relation extends to lower mass ranges. Using the velocity dispersion values of $61.2\pm10.1$ km s$^{-1}$ \citep{ho2009} for NGC 3367 and $85\pm1$ km s$^{-1}$ \citep{batcheldor2005} for NGC 4536 and the M-$\sigma$ relation from \citet{gultekin2009b}, the masses are found to be consistent with the lower mass limits derived using the Eddington limit, but significantly higher than those derived using the BMC model.

{\it Fundamental Plane} --- This method of estimating black hole mass uses an apparent correlation between X-ray luminosity, 5GHz radio luminosity, and black hole mass. \citet{gultekin2009a} recently examined this correlation, using dynamically-determined black hole masses, archival data from Chandra, and 5GHz luminosities from the literature for 18 galaxies. Using the X-ray luminosity from this paper and the 5GHz core radio flux estimated earlier, mass estimates for our galaxies may be determined. The masses estimated with this method are consistent with masses implied by the M-$\sigma$ relation as can be seen in Table 3. The relation however comes with in intrinsic scatter 0.77 dex, larger than the scatter of the M-$\sigma$ relation, but like the M-$\sigma$ relation, the scatter increases as the mass decreases. The relation is developed using higher mass black hole on the order of 10$^{7}$ to 10$^{9}$ $M_{\odot}$ and may not be appropriate for this sample if their central black holes truly have relatively small masses. \citet{gultekin2009a} explains this scatter with the possibility that the relation may not apply to sources that accrete at high rates. They also explain that the scatter may be skewed by a few outliers.

\begin{table}
\begin{center}
\scriptsize
\begin{tabular}{lccccc}
\multicolumn{6}{l}{{\bf Table 3: Black Hole Mass Estimates}}\\
\hline
\hline
\noalign{\smallskip}
     Galaxy  &  log$M_{\rm Edd}$  & log$M_{\rm Edd}$ &  log$M_{\rm BMC}$ & log$M_{\rm \sigma}$   &  log$M_{\rm fund}$ \\   
&[NeV]&X-ray&&&\\ 
\noalign{\smallskip}
\hline
\noalign{\smallskip}
  NGC~3367       & $>$5.19   &  $>$3.93    &  3.85 - 5.40     & 5.94  & 6.85 - 6.92   \\
\noalign{\smallskip}
\hline
\noalign{\smallskip}
   NGC~4536      & $>$3.91   &  $>$3.58    &  $<$4.15            &  6.52  &  $<$6.67    \\ 
\noalign{\smallskip} 
\hline
\hline
\end{tabular}
\end{center}
\footnotesize
\label{tab3}
\end{table}

We note that the last two methods yield significantly larger values for the black hole than the ones obtained using the scaling technique based on the results from the BMC fit. However, \citet{gliozzi2010} using very high-quality X-ray spectra of the Narrow line Seyfert 1 galaxy PKS 0558-504, showed that the value of $M_{\rm BH}$ obtained with this technique is fully consistent with the other indirect methods. The low values of $M_{\rm BH}$ derived for NGC 3367 and NGC 4536 using the scaling technique can be reconciled with those derived from the fundamental plane and from the M-$\sigma$ relation if the measured X-ray luminosity is substantially underestimated (for example because of a significant intrinsic absorption). This scenario is consistent with the as the anomalous optical - X-ray spectral index discussed in section 3.3.

\section{Summary and Conclusions}
 
We have analyzed the {\it XMM-Newton} observations of two late-type galaxies; NGC 3367 and NGC 4536, which are believed to harbor AGNs based on the previous detection of mid-infrared [NeV] line emission.  Our main results are summarized as follows:

\begin{enumerate} 

\item Detailed spectral analysis of the {\it XMM-Newton} data for these galaxies reveals that the X-ray luminosity is dominated by a power law  with 2-10 keV luminosities of $2.0\times 10^{40}$ ergs s$^{-1}$ for NGC 3367 and $0.9\times 10^{40}$ ergs s$^{-1}$ for NGC 4536, respectively, consistent with low luminosity AGNs.

\item The possibility that significant X-ray emission may be emitted by X-ray binaries was explored. It was found that XRBs could be responsible for only a small fraction of the X-ray luminosity detected, and that the significant [NeV] emission cannot be explained by the presence of XRBs alone.  

\item Low-amplitude flux changes may be present in both galaxies, and NGC 4536 shows statistically significant variability, providing further evidence of an AGN.

\item A comparison of the black hole mass estimated using the BMC model with other methods suggests the X-ray source may be absorbed in both galaxies. This hypothesis is supported by the steep spectral index $\alpha_{OX}$ found in both sources.   

\item Estimated black hole masses range of 10$^{5}$ - 10$^{7}$ $M_{\odot}$ for NGC 3367 and 10$^{4}$ - 10$^{6}$ $M_{\odot}$ for NGC 4536. 

\end{enumerate}

We have demonstrated that MIR spectroscopy coupled with X-ray observations can more robustly determine the presence of an AGN and estimate the black hole mass.

\acknowledgements

It is a pleasure to thank Marla Katz for her invaluable help in the initial data analysis required for this project. This work would not have been possible without her dedication and hard work.  We are also very thankful for the helpful comments from the referee, which improved this paper. This research has made use of the NASA/IPAC Extragalactic Database (NED), which is operated by the Jet Propulsion Laboratory, California Institute of Technology, under contract with the National Aeronautics and Space Administration.  S.S. gratefully acknowledges support by the XMM-Newton Guest Investigator Program under NASA grant NNX08AZ39G.  SS and RMS gratefully acknowledge funds from NASA grant NAG5-1078.

\end{document}